\documentclass[useAMS,usenatbib,referee]{mn2e}
\usepackage{}

\usepackage{txfonts}

\usepackage{natbib}  
\usepackage{graphicx}
\usepackage{txfonts}
\usepackage{mathrsfs}

\title[Asteroseismic analysis of two $\alpha$-enhanced stars KIC 7976303 and KIC 8694723]{Asteroseismic analysis of two $\alpha$-enhanced stars KIC 7976303 and KIC 8694723}
\author[Z. S. Ge et al.]{Z. S. Ge$^{1}$\thanks{E-mail:gezhishuai@mail.bnu.edu.cn}, S. L. Bi$^{1}$\thanks{E-mail:bisl@bnu.edu.cn}, T. D. Li$^{2}$, K. Liu$^{1}$, Z. J. Tian$^{1}$, W. M. Yang$^{1}$, Z. E. Liu$^{1}$ and J. Yu$^{1}$
\\
$^{1}$Department of Astronomy, Beijing Normal University, Beijing 100875, China\\
$^{2}$ Key Laboratory of Solar Activity, National Astronomical Observatories, Chinese Academy of Science, Beijing 100012, China}

\begin{document}

\pagerange{\pageref{firstpage}--\pageref{lastpage}} \pubyear{2002}

\maketitle

\label{firstpage}

\begin{abstract}

We intent to use stellar models with and without $\alpha$-enhancement, as well as asteroseismic analysis, to study two $\alpha$-enhanced stars, KIC 7976303 and KIC 8694723. For the $\alpha$-enhanced models, we adopt [$\alpha$/Fe] = 0.2, and 0.4. For the evolved sub-giant KIC 7976303 with mixed-modes, $\alpha$-enhanced models fit the observations better than the models without $\alpha$-enhancement, and point to a star with $M$ = 1.20 $\pm$ 0.02 $M_\odot$, $t$ = 4.88 $\pm$ 0.08 Gyr, $R$ = 2.04 $\pm$ 0.01 $R_\odot$, and $L$ = 5.07 $\pm$ 0.05 $L_\odot$. For the post turn-off star KIC 8694723, we find that the models fit the observations well in all three cases ([$\alpha$/Fe] = 0.0, 0.2, 0.4). According to the observed metal abundances, only $\alpha$-enhanced models are used to estimate the stellar parameters, which are $M$ = 1.13 $\pm$ 0.06 $M_\odot$, $t$ = 5.47 $\pm$ 0.43 Gyr, $R$ = 1.55 $\pm$ 0.13 $R_\odot$, and $L$ = 3.18 $\pm$ 0.26 $L_\odot$. Our $\alpha$-enhanced models indicate KIC 7976303 has a larger mass and younger age than in previous works based on standard models. Furthermore, the differences of estimated mass and age between the three cases are $\sim$0.1 $M_\odot$ and $\sim$ 0.5 - 1.3 Gyr. These results suggest that we include $\alpha$-enhancement in the modeling of $\alpha$-enhanced stars, such as members of GCs (Globular Clusters), metal-poor stars in the disc and in the halo.

\end{abstract}

\section{Introduction}

Metallicity is an important property of stars. The chemical composition of the Sun has been estimated, and has been updates several times in the past decades \citep{GN93,GS98,AGS05,AGSS09}. The chemical composition of the Sun, which is taken as standard, has been set as [Fe/H]$_\odot$ = 0.0 and [$M$/Fe]$_\odot$ = 0.0 (where $M$ denotes metal element). In stellar modeling, we generally assume that for all stars, whether they are metal-poor ([Fe/H] $<$ 0.0) or metal-rich ([Fe/H] $>$ 0.0), their metal- element mixtures are the same as the Sun, that is to say, [M/Fe] = 0.0. However, the scaled-solar metal mixture is not universal. One pattern presented in \citet{Greenstein1970} is formed with enhanced abundances of $\alpha$-capture elements (i.e. O, Ne, Mg, Si, S, Ar, Ca, Ti). In other words, the number abundance ratio [$\alpha$/Fe], expressed as a fraction of the solar value, is greater than unity (i.e. [$\alpha$/Fe] $>$ 0.0).

Observations show that most metal-poor stars in the solar neighborhood and in globular clusters (GCs) are $\alpha$-enhanced and their metal abundance ratios are significantly different from those of the Sun and the metal-rich stars in the Galactic Disc \citep{Wheeler1989}. With more recent observations, it is confirmed that most metal-poor field stars, whether they are in the halo \citep{Nissen2010,Nissen2011,Schuster2012} or in the disc \citep{Bensby2003,Bensby2005,Bensby2007}, are $\alpha$-enhanced. For the disc stars in the metallicity range of -1.4 $<$ [Fe/H] $<$ -0.7, the value of [$\alpha$/Fe] is about 0.1 $\sim$ 0.4, declining with increasing metallicity. For the halo stars, similar trends can be found, and stars belonging to inner-halo generally have higher values of [$\alpha$/Fe] than those of outer-halo ones \citep{Bensby2003}. For the case of GCs in the Galactic Halo, [$\alpha$/Fe] stays constant ($\simeq$ 0.3) in stars of [Fe/H] $<$ -1.0, declines as metallicity increases in the range of -1.0 $<$ [Fe/H] $<$ 0.0, reaches 0.0 at [Fe/H] $\sim$ 0.0 \citep{Muciarelli2013}.

Because an $\alpha$-enhanced mixture is one of the most interesting patterns which significantly changes the opacities of a star, it also has an influence on the evolution and structure of a star. The role played by $\alpha$-enhancement in evolutions and isochrones is well understood \citep{VandenBerg2000a,Kim2002,Pietrinferni2006,Dotter2007,Coelho2007}. Stellar tracks and isochrones of $\alpha$-enhanced mixtures have hotter/bluer turnoffs and red giant branches in the H-R diagram (Hertzsprung - Russell diagram) than those computed with scaled-solar mixtures  \citep{Salaris1993,SW98,VandenBerg2000a,Salasnich2000}. New isochrones that take $\alpha$-enhancement into account have also been widely used for estimating the ages of GCs and old field stars. The ages determined using the updated models are younger than those determined by previous works \citep{VandenBerg2000b,Bergbusch2001,Yi2001,Kim2002}. These works suggest that it is necessary to include $\alpha$-enhancement in related studies.

KIC 7976303 and KIC 8694723 are two targets of the $Kepler$ mission, which are found to present $\alpha$-enhanced features in metal mixtures. \citet{Bruntt2012} have provided detailed metal abundances of the two $Kepler$ stars. \footnote{http://vizier.cfa.harvard.edu/viz-bin/VizieR?-source=J/MNRAS/423/122} It is found that the relative abundances of several $\alpha$-elements (O, Mg, Si, Ca, Ti) are higher than that of the Sun, the detailed metal mixtures can be found in Table \ref{Tab:7976303mix} and \ref{Tab:8694723mix} in Appendix. KIC 7976303 and KIC 8694723 are stars with the spectrum type of G0 and F5. The effective temperatures of the two stars are 6005 $\pm$ 110 K and 6101 $\pm$ 110 K, and the values of $\log$ g are 4.23 $\pm$ 0.5 dex and 4.15 $\pm$ 0.5 dex, as given by the $Kepler$ Input Catalog \citep{Brown2011}. The effective temperatures were then revised by \citet{Pinsonneault2012} using Sloan Digital Sky Survey (SDSS) $griz$ filters; the results are 6260 $\pm$ 51 K and 6310 $\pm$ 56 K. \citet{Bruntt2012} carried out a detailed spectroscopic study of these stars and found the temperatures of two stars to be 6095 $\pm$ 70 K and 6200 $\pm$ 70 K. The values of log $g$ also have been revised by asteroseismic analysis in this work; the results are 3.87 $\pm$ 0.03 and 4.10 $\pm$ 0.03 separately. These two stars are metal-poor. The values of metallicity for KIC 7976303 and KIC 8694723 are -0.59 and -0.51, as provided by $Kepler$ Input Catalog, with an uncertainty of 0.5. The values of [Fe/H] are corrected by asteroseismic $\log$ g, leading to the results of -0.53 $\pm$ 0.06 and -0.59 $\pm$ 0.06 \citep{Bruntt2012}.

\citet{Appourchaux2012} obtained oscillation frequencies of KIC 7976303 and KIC 8694723 by analyzing the power spectra using both maximum likelihood estimators and Bayesian estimators. Individual frequencies are provided with small errors because the data they employed are a 9-months time series. The figure of large frequency separation as a function of effective temperature (their fig.1) showed that both stars have left the main-sequence. Moreover, KIC 7976303 has been studied by \citet{Mathur2012} using oscillation analysis and modelling. First, they extracted oscillation frequencies with one month's observation data from the $Kepler$ satellite, and then estimated stellar parameters of the star with four stellar modelling methods. They obtained the mass and the age of the star are 1.01-1.19 $M_\odot$ and 4.7-7.0 Gyr. They noticed that the results of KIC 7976303 cannot fit both atmospheric constraints and individual frequencies. These deviations were thought to be from spectra observations, abundance analysis or inadequate modelling.

In this work, we intend to determine the stellar parameters of KIC 7976303 and KIC 8694723 using $\alpha$-enhanced stellar models. In Section 2, we present observation constraints for these two stars. In Section 3, we introduce our stellar models. We present the asteroseismic diagnostics and the parameters determined for the two stars in Section 4. Finally, we summarize our results and give conclusions in Section 5.

\section{Observation constraints}

Table \ref{Tab:obs} presents the observations of these two stars, which  have been carried out in several works \citep{Brown2011,Pinsonneault2012,Bruntt2012,Appourchaux2012}. We use the detailed metal abundances provided by \citet{Bruntt2012}, to keep consistency, we employ atmospheric parameters ($T_{\rm{eff}}$, $\log$ g and [Fe/H]) in the following analysis. \citet{Bruntt2012} have presented a detailed spectroscopic study by analysing high-quality spectra adopted from two service observing programs during 2010 May-September \citep{Bruntt2012} using the ESPaDOnS spectrograph at 3.6-m Canada-France-Hawaii Telescope (CFHT) in the United States \citep{Donati2006} and the NARVAL spectrograph mounted on the 2-m Bernard Lyot Telescope at the Pic du Midi Observatory in France. Because of the high-quality spectra, several abundances of $\alpha$-elements were obtained, and the data can be seen in Table \ref{Tab:7976303mix} and \ref{Tab:8694723mix}. Following the calculation process presented in Appendix, we obtain values of [$\alpha$/Fe] = 0.36 and 0.35 for KIC 7976303 and KIC 8694723, respectively. The luminosities of KIC 7976303 and KIC 8694723 have not been provided in previous works. Hence, we use the surface gravity (log $g$) and effective temperature ($T_{\rm{eff}}$) to construct $T_{\rm{eff}}$-log $g$ diagrams in order to check the evolution stage of the two stars.

Asteroseismic frequencies are adopted from \citet{Appourchaux2012}, because they use data of a nine-month time series and analyse the power spectra using both maximum likelihood estimators and Bayesian estimators. The values of $\langle \Delta\nu \rangle$ they have provided are the median values of $\Delta\nu $, while the values we have adopted are calculated from individual frequencies by fitting a straight line to the frequencies and radial orders ($n$). The slope of the fitted line is the value of the mean large frequency separation. The uncertainty of the mean large frequency separation is then the uncertainty of the slope, which is calculated by linear regression. In this work we use the 3$\sigma$ error as the observation uncertainty of $\langle \Delta\nu \rangle$.

\begin{table*}
 \centering
 \begin{minipage}{140mm}
  \caption{Observation Constraints}             
  \label{Tab:obs}
   \begin{tabular}{l c c c c c c  }
\hline\hline                 
     KIC     & $T_{\rm{eff}}$ & $\log$ g  & [Fe/H]    & $\langle\Delta\nu\rangle$ & $\nu_{\rm{max}}$ &  Ref. \\
             & (K)  &  &   (dex)   & ($\mu$Hz)   & ($\mu$Hz)   &        \\  
\hline
      7976303         & 6260 $\pm$ 51      & ...              & ...                  & ...               & ...      &  1  \\      
                      & 6095 $\pm$ 70      & 3.87 $\pm$ 0.03  & -0.53 $\pm$ 0.06      & ...               & ...      &  2 \\
                      & 6005 $\pm$ 110     & 4.23 $\pm$ 0.5   & -0.59 $\pm$ 0.5       & ...               & ...      &  3  \\
                      & ...               & ...              & ...                  & 51.3  \footnote{$\langle\Delta\nu\rangle$ derived from the $\ell$=0 modes, for KIC 7976303 is with mixed-modes.\\1. \citet{Pinsonneault2012}\\2. \citet{Bruntt2012}\\3. KIC \citep{Brown2011}\\4. \citet{Appourchaux2012}\\5. \citet{Mathur2012}\\6.The photometric data we used in this work come from \citet{Bruntt2012}. The values of large frequency separation and the uncertainties come from individual frequencies provided by \citet{Appourchaux2012}. }     & 826       &  4 \\
                      & ...               & ...              & ...                 & 50.95 $\pm$ 0.37   & 910$\pm$ 25 &  5  \\
                      & 6095 $\pm$ 70     & 3.87 $\pm$ 0.03  & -0.53 $\pm$ 0.06    & 51.20 $\pm$ 0.40   & 826 $\pm$ 55 &   6   \\
  \hline                                   
      8694723         & 6310 $\pm$ 56      & ...              & ...                  & ...               & ...      &  1  \\
                      & 6200 $\pm$ 70      & 4.10 $\pm$ 0.03  & -0.59 $\pm$ 0.06     & ...               & ...      &  2   \\
                      & 6101 $\pm$ 110     & 4.15 $\pm$ 0.5   & -0.51 $\pm$ 0.5      & ...               & ...      &  3   \\
                      & ...                & ...              & ...                  & 75.1              & 1384     &  4  \\
                      &  6200 $\pm$ 70       &  4.10 $\pm$ 0.03   &  -0.59 $\pm$ 0.06     & 74.54 $\pm$ 0.47     &  1384 $\pm$ 92     &  6  \\
  \hline                                   
\end{tabular}
\end{minipage}
\end{table*}

\section{Stellar Models}

\subsection{Input Physics}

We compute a grid of evolutionary tracks using the Yale Rotation and Evolution Code \citep{Guenther1992}, in order to estimate parameters of these two $Kepler$ stars. The helium abundance is set to be constant ($Y$ = 0.248), which is the standard big bang nucleosynthesis value \citep{Spergel2007}. The mixing-length parameter $\alpha_{\ell}$ is a solar calibrated value, 1.75. For standard models, we use the scaled-solar mixture of \citet{GS98}. In $\alpha$-enhanced models, the abundances of all $\alpha$-elements (i.e. O, Ne, Mg, Si, S, Ca and Ti; Ar is ignored because it is an inertial gas with very low abundance) have been increased by 0.2 and 0.4 dex. We use OPAL high-temperature opacity tables \footnote{http://opalopacity.llnl.gov/new.html} along with \citet{Ferguson2005} opacities for low temperatures. The models are calculated using the updated OPAL equation-of-state tables EOS2005 \citep{Rogers2002}. All models include gravitational settling of helium and heavy elements using the formulation of \citet{Thoul1994}.

For Population I stars, the relationship between [Fe/H] and ratio of surface metal-element abundance to hydrogen abundance ($Z/X$) is $\log$ ($Z/X$) = $\log$ ($Z/X$)$_\odot$ + [Fe/H], where ($Z/X$)$_\odot$ is the ratio of the metal element to hydrogen for scaled-solar mixture, which is regarded as 0.023 for the metal mixture of \citet{GS98}. However, this relationship is not adequate for $\alpha$-enhanced mixtures. For a certain [Fe/H], an $\alpha$-enhanced mixture leads to a larger $Z/X$ than the scaled-solar mixture, which requires us to use detailed element abundances to calculate $Z/X$. The description of the calculation can be checked in Appendix. Table \ref{Tab:ZX} lists the quantities of $Z/X$ for a certain [Fe/H] with different [$\alpha$/Fe].

\begin{table}
 \centering
\renewcommand\arraystretch{1.0}
\caption[]{ Metallicity ralation.}
\label{Tab:ZX}
\begin{tabular}{cccc}
\hline\hline                 
      $Z/X$                      &          & $Z/X$                      & $Z/X$         \\
      $[\alpha /{\rm{Fe}}]$ = 0.0 & [Fe/H]   & $[\alpha /{\rm{Fe}}]$ = 0.2 & $[\alpha /{\rm{Fe}}]$ = 0.4 \\
      \hline
       0.0078   & -0.47   & 0.0110   & 0.0159    \\
       0.0068   & -0.53   & 0.0095   & 0.0139    \\
       0.0059   & -0.59   & 0.0083   & 0.0121    \\
       0.0051   & -0.65   & 0.0072   & 0.0105    \\
    \hline                                   
       \end{tabular}
       \end{table}

\begin{table}
 \centering
\renewcommand\arraystretch{1.0}
\caption[]{ Input parameters.}
\label{Tab:input}
 \begin{tabular}{ccc}
 \hline\hline                 
                  KIC 7976303           & range        & $\delta$  \\
                  $M$ $(M_{\odot})$        & 0.84-1.26    & 0.02     \\
                  $Z_{i}$               & 0.007-0.018  & 0.001    \\

$[\alpha /{\rm{Fe}}]$                   & 0.0-0.4      & 0.2      \\
                  $Y_{i}$               & 0.248 & ...   \\
                  $\alpha$              & 1.75  & ...   \\
\hline               
                  KIC 8694723           & range        & $\delta$ \\
                  $M$ $(M_{\odot})$        & 0.90-1.32    & 0.02      \\
                  $Z_{i}$               & 0.006-0.017  & 0.001     \\
                 $[\alpha /{\rm{Fe}}]$  & 0.0-0.4      &0.2        \\
                  $Y_{i}$               & 0.248 & ...  \\
                  $\alpha$              & 1.75  & ...  \\
\hline             
\end{tabular}
\end{table}

The mass ranges of the models are predicted by observed properties, that is effective temperature $T_{\rm{eff}}$, $\nu_{\max}$, and the  large frequency separation $\langle \Delta\nu \rangle$. \citet{Brown1991} put a scaling relation to predict $\nu_{\rm{max}}$,
\begin{equation}
\
\frac{{\nu _{\max } }}{{\nu _{\max , \odot } }} \approx \left( {\frac{M}{{M_ \odot  }}} \right)\left( {\frac{R}{{R_ \odot  }}} \right)^{ - 2} \left( {\frac{{T_{{\rm{eff}}} }}{{T_{{\rm{eff}}, \odot } }}} \right)^{ - 1/2}.
\label{vmax}
\end{equation} \citet{Kjeldsen1995} give the relation to estimate the mean large frequency separation ($\langle \Delta\nu \rangle$),
\begin{equation}
\frac{{\langle \Delta\nu \rangle}}{{\langle \Delta\nu \rangle _ \odot  }} \approx \left( {\frac{M}{{M_ \odot  }}} \right)^{1/2} \left( {\frac{R}{{R_ \odot  }}} \right)^{ - 3/2}.
  \label{deltnu}
\end{equation} With Equation (\ref{vmax}) and Equation (\ref{deltnu}), the stellar mass can be estimated by
\begin{equation}
\frac{M}{{M_ \odot  }} \approx \left( {\frac{{\langle \Delta\nu \rangle}}{{\langle \Delta\nu \rangle _ \odot  }}} \right)^{ - 4} \left( {\frac{{\nu _{\max } }}{{\nu _{\max , \odot } }}} \right)^3 \left( {\frac{{T_{{\rm{eff}}} }}{{T_{{\rm{eff}}, \odot } }}} \right)^{3/2}.
  \label{mass}
\end{equation}
The uncertainty of the predicted mass is from error propagation formula. Using the relationship between $Z/X$ and [Fe/H] presented in Table \ref{Tab:ZX}, we estimate the initial metal abundances, $Z_{i}$. The input parameters for stellar models are listed in Table \ref{Tab:input}.

\subsection{Model Calibration}

There are a large amount of models falling within the error box in the $T_{\rm{eff}}$-$\log$ g diagram, which fit $T_{\rm{eff}}$ and $\log$ g. For each model in the error box, we calculated the adiabatic low-$\ell$ $p$-mode frequencies using the pulsation code of \citet{Guenther1994}. Considering the observational individual frequencies, the frequencies of $\ell$ = 0, 1, 2 are calculated for a radial order n ranging from 10 to 20 in the frequency range of 500-1200 $\mu$Hz for KIC 7976303. For KIC 8694723, frequencies of $\ell$ = 0, 1, 2 are calculated for a radial order $n$ ranging from 10 to 20 in the frequency range of 700-2000 $\mu$Hz.

Considering the surface effect on $p$-mode frequencies, we employ the method by \citet{Kjeldsen2008} to make corrections on the theoretical frequencies of the selected models. The correction equation is
 \begin{equation}
\nu _{{\rm{corr}}}  - \nu _{\bmod }  = a_0 \left( {\frac{{\nu _{\bmod } }}{{\nu _{\max } }}} \right)^b,
\end{equation} where b is fixed to a solar calibrated value as 4.99, $a_0$ is the size of correction at $\nu_{\max}$. The value of $a_0$ for each model is calculated by the equation below \citep{Kjeldsen2008},
\begin{equation}
a = \frac{{\left\langle {\nu _{{\rm{obs}}} \left( n \right)} \right\rangle  - r\left\langle {\nu _{\bmod } \left( n \right)} \right\rangle }}{{N^{ - 1} \sum\nolimits_{i = 1}^N {\left[ {\nu _{{\rm{obs}}} \left( {n_i } \right)/\nu _0 } \right]^b } }}.
\end{equation}

\begin{figure*}
\centering
\includegraphics[width=16.0cm]{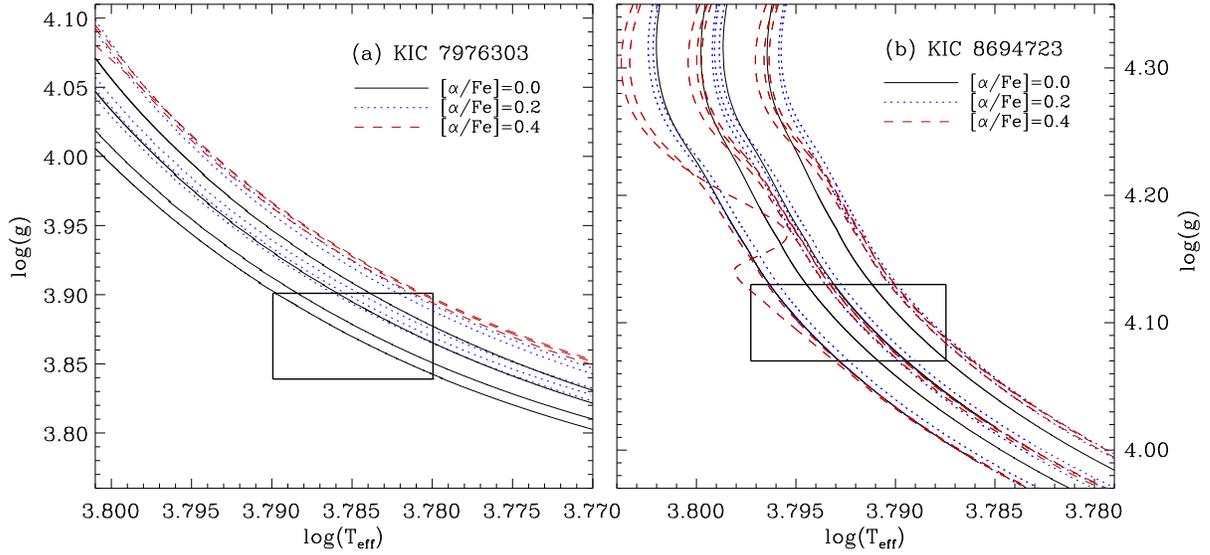}
   \caption[]{$T_{\rm{eff}}$ - $\log$ g diagrams for KIC 7976303 (panel (a)) and KIC 8694723 (panel (b)). The black lines represent tracks with [$\alpha$/Fe] = 0.0, the blue dotted ones are tracks with [$\alpha$/Fe] = 0.2, the red dashed lines represent tracks with [$\alpha$/Fe] = 0.4. All the tracks fit spectroscopic constraints and observed $\langle \Delta\nu \rangle$.}
\label{fig:HR}
\end{figure*}

After the surface correction, the theoretical $\langle \Delta\nu \rangle$ values can be calculated from individual frequencies of the models by fitting a straight line to frequencies $\nu_{n,l}$ and n within the observation frequency range. We further use observed [Fe/H] (which is replaced by $Z/X$) and $\langle \Delta\nu \rangle$ to select models. Hence, we obtain models that are within observation constraints ( $T_{\rm{eff}}$, $\log$ g, [Fe/H], and $\langle \Delta\nu \rangle$). The corresponding evolutionary tracks in the $T_{\rm{eff}}$-$\log$ $g$ diagrams in Fig. \ref{fig:HR} show that KIC 7976303 (Fig. \ref{fig:HR}a) is an evolved sub-giant star and that KIC 8694723 (Fig. \ref{fig:HR}b) is a post turn-off star.

To examine the agreement between the theoretical and observational frequencies, we employ the following function:
\begin{equation}
\chi _\nu ^2  = \frac{1}{N}\sum\limits_{n,l}^{} {\left( {\frac{{\nu _l^{theo} (n) - \nu _l^{obs} (n)}}{{\sigma \nu _l^{obs} (n)}}} \right)^2 },\
  \label{XV}
\end{equation}
Here, $N$ is the total number of modes, $\nu _l^{theo} (n)$ and $\nu _l^{obs} (n)$ represent the corrected theoretical and observed frequencies, respectively, and $\sigma \nu _l^{obs} (n)$ is the error for the individual observational frequencies.

As described in equation (\ref{XV}), $\chi_{\nu}^2$ are used to seek the best-fitting models. Because the errors of observed individual frequencies are very small, $\chi_{\nu}^2$ is very sensitive to the deviations between observed and theoretical frequencies, and thus the results tend to be numerically large. It is generally accepted that models with lower values of $\chi_{\nu}^2$ might fit better with observations. Hence, we select models with $\chi_{\nu}^2$ $\leq$ 100 as candidates to find the best-fitting models. It is noticed that dipole modes with mixed-modes are not included in the calculation; that is to say, only frequencies of $\ell$ = 0 and 2 are used to calculate $\chi_{\nu}^2$ values for KIC 7976303.

\section{Asteroseismic Diagnostic}

\subsection{KIC 7976303}

\subsubsection{The \'Echelle Diagrams}

\begin{figure*}
\centering
\includegraphics[width=16.0cm]{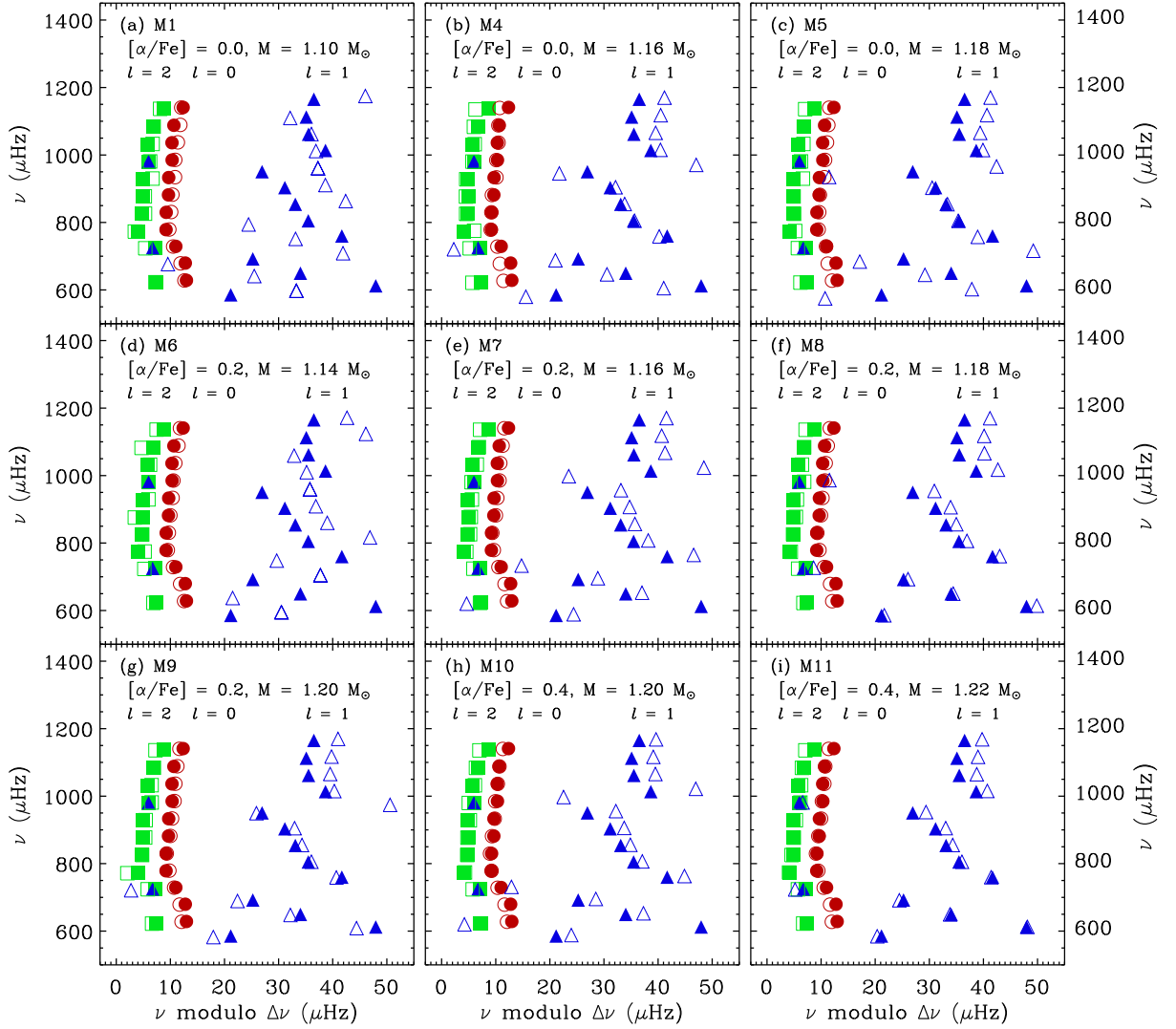}
\caption[]{\'Echelle diagrams for all the selected models (M1, M4, M5, M6, M7, M8, M9, M10, and M11) of KIC 7976303. The filled signs represent observations, the hollow ones represent data from models. Red circles represent frequencies for $\ell$ = 0 modes, blue triangles means frequencies for $\ell$ = 1 modes, green squares means modes for $\ell$ = 2 modes. The first row shows models with [$\alpha$/Fe] = 0.0. The middle two rows present models with [$\alpha$/Fe] = 0.2. The bottom row presents models with [$\alpha$/Fe] = 0.4. $\langle\Delta\nu\rangle_{\ell=0}$ = 51.20 $\mu$Hz.}
\label{fig:7976303ED}
\end{figure*}

The models with $\chi _{\nu _{\ell {\rm{ = }}0,2} }^2$ $\leq$ 100 are presented in Table \ref{Tab:7976303} as candidates for selecting the best-fitting models. To use individual frequencies as constraints, we analyse the behaviours of the oscillation frequencies of these models. The echelle diagrams of models that have a relatively lower $\chi _{\nu _{\ell {\rm{ = }}0,2} }^2$ value are presented in Fig. \ref{fig:7976303ED}. From this figure, we find that these models fit observations well for $\ell$ = 0 and $\ell$ = 2 modes.
It is generally acknowledged that oscillation frequencies are very sensitive to the stellar interior structures. The structures described by the models, which have relatively lower $\chi _{\nu _{\ell {\rm{ = }}0,2} }^2$ values, might be close to the star. However, to estimate the parameters of KIC 7976303, we are required to find models that reproduce behaviors of all the frequencies well, including mixed-modes.

\begin{table*}
 \centering
 \begin{minipage}{200mm}
\caption[]{Theoretical stellar parameters from models for KIC 7976303.}
\label{Tab:7976303}
 \begin{tabular}{ccccccccccccccc}
 \hline\hline                 
    models& $M$         &	$Z_{i}$ & [$\alpha$/Fe] & $t$        & $L$          & $R$    & $\log$ g & $T_{\rm{eff}}$  & $Z/X$ & [Fe/H]  & $\langle\Delta\nu\rangle$ &     $\chi _{\nu _{\ell {\rm{ = }}0,2} }^2$  \\
          &($M_{\odot}$)&           &          &  (Gyr)     &  ($L_{\odot}$)  & ($R_{\odot}$)              &       & (K)        &             & (dex) &         &   ($\mu$Hz)       & \\
 \hline\noalign{\smallskip}
 1 & 1.10  &  0.007  &  0.0  &  5.784  & 4.72 &  1.99 & 3.88  & 6041.7   &  0.007  &-0.54   & 51.32   & 17    \\
 2 &1.14 & 0.008 & 0.0 & 5.280 & 4.95 &  2.00 & 3.89 & 6089.2 & 0.007 &-0.50 & 51.37   &	 53    \\
 3 &1.14 & 0.008 & 0.0 & 5.285 & 4.96 &  2.00 & 3.89 & 6084.3 & 0.007 &-0.50 & 51.21   &	 70    \\

 4 &1.16 & 0.008 &  0.0 &  4.872  &  5.20 &   2.01  &  3.90  &  6154.5  &  0.006  & -0.57 & 51.33   & 20    \\

 5 &1.18 & 0.009 & 0.0 & 4.810 & 5.21 &  2.02 & 3.90 & 6138.5 & 0.007 &-0.50 & 51.32   &	 17    \\

 \hline\noalign{\smallskip}
 6 & 1.14  &  0.009  &  0.2  &  5.474  & 4.78  &  2.01 &  3.89  &  6025.0  &  0.009  &  -0.56  & 51.29  & 14  \\

 7 &1.16 & 0.009 & 0.2 & 5.087 & 5.04 & 2.02 & 3.89 & 6095.4 & 0.008 &-0.59 & 51.30  &  9    \\

 $\textbf{8}$ & $\textbf{1.18}$ & $\textbf{0.010}$ & $\textbf{0.2}$ & $\textbf{4.958}$ & $\textbf{5.08}$ & $\textbf{2.03}$ & $\textbf{3.90}$ & $\textbf{6090.9}$ & $\textbf{0.009}$ & $\textbf{-0.54}$ & $\textbf{51.30}$    & $\textbf{8}$  \\

  $\textbf{9}$& $\textbf{1.20}$ & $\textbf{0.011}$ & $\textbf{0.2}$ & $\textbf{4.830}$ & $\textbf{5.11}$ & $\textbf{2.04}$ & $\textbf{3.90}$ & $\textbf{6083.0}$ & $\textbf{0.010}$ & $\textbf{-0.50}$ & $\textbf{51.30}$    & $\textbf{13}$  \\

 \hline\noalign{\smallskip}
 10 & 1.20 &  0.012 & 0.4 &  4.952 & 4.96 & 2.04 & 3.90  &  6030.9  & 0.012  & -0.58 &  51.25   &  7  \\

 $\textbf{11}$ & $\textbf{1.22}$ & $\textbf{0.013}$ & $\textbf{0.4}$ & $\textbf{4.798}$ & $\textbf{5.02}$ & $\textbf{2.05}$ & $\textbf{3.90}$ & $\textbf{6030.5}$ & $\textbf{0.013}$ & $\textbf{-0.56}$  &  $\textbf{51.27}$  &	 $\textbf{7}$   \\

 12 &1.24 & 0.014 & 0.4 & 4.654 &5.08 & 2.07 & 3.90 & 6031.7 & 0.014 &-0.52 &  51.18  &	 59    \\

 \hline              
\end{tabular}
\end{minipage}
\end{table*}

\begin{figure*}
\centering
\includegraphics[width=16.0cm]{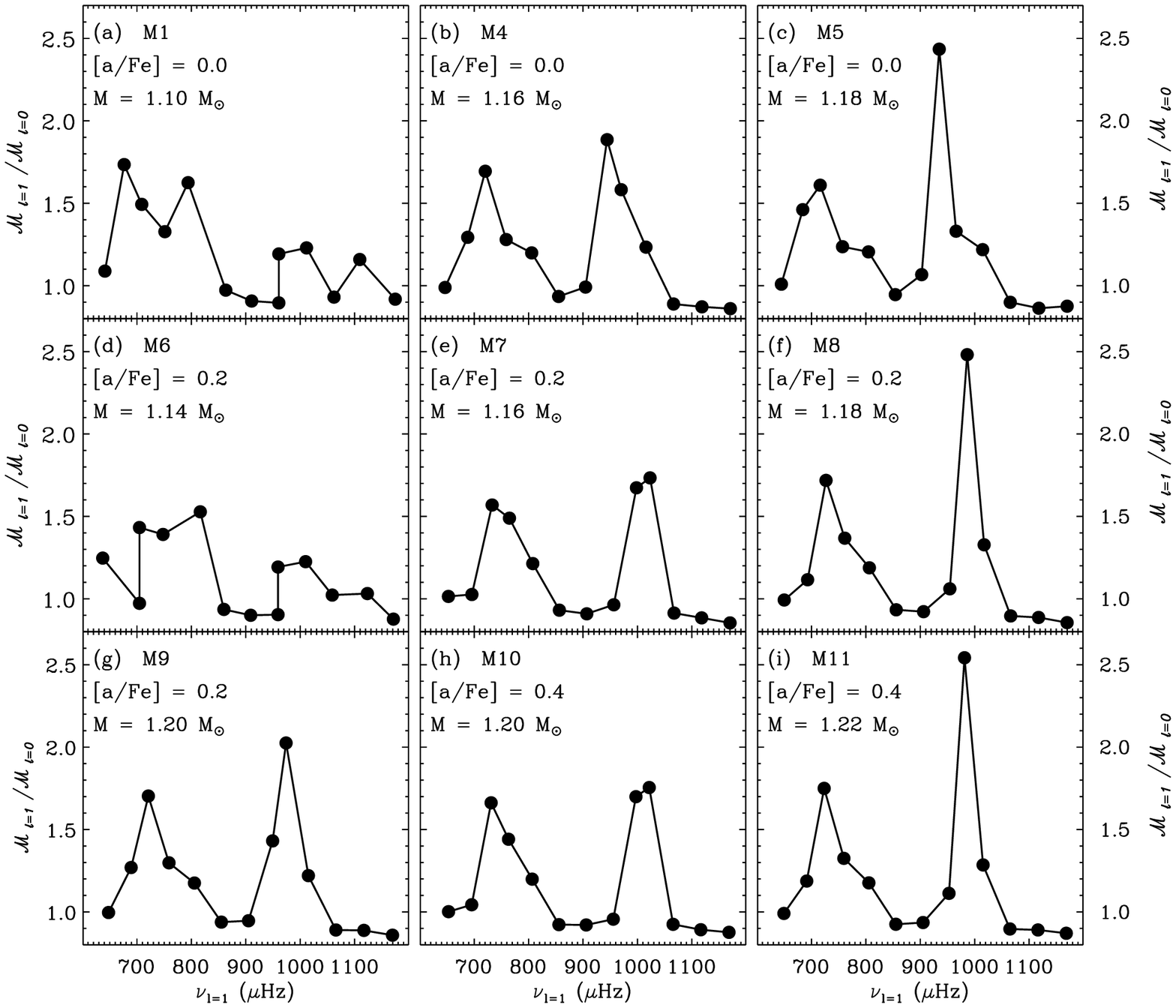}
\caption[]{Mode-mass ratio as a function of frequencies for models M1, M4, M5, M6, M7, M8, M9, M10, and M11 of KIC 7976303. The up row shows models with [$\alpha$/Fe] = 0.0. The middle two rows present models with [$\alpha$/Fe] = 0.2. The bottom row presents models with [$\alpha$/Fe] = 0.4.}
\label{fig:7976303mass}
\end{figure*}

\subsubsection{The Mixed-mode}

The interactions between the $p$-mode and $g$-mode cavities lead to many avoided crossings. Avoided crossings occur on a very short time-scale compared with the stellar evolution time-scale, which causes inherent difficulties in finding best-fitting models \citep{Deheuvels2011}. The behaviours of avoided crossing of $\ell$ = 1 modes are related to coupling strength. The coupling strength of the mixed-mode corresponds to the ratios of mode inertia\citep{Benomar2014}, which are proportional to the mode-mass ratio. The mixed-modes closest to the $\ell$ = 1 $p$-mode frequencies have the lowest mode-mass ratios, while those with higher mode-mass ratios stand for modes that depart significantly from the $p$-mode \citep{Benomar2014}. Therefore, the mode-mass ratio can be used as a diagnostic of the behaviors of avoided crossing.

Previously, we have obtained models that fit $\ell$ = 0 and $\ell$ = 2 modes well. To describe the characteristics of mixed-modes, we use both echelle diagrams and mode-mass ratios ($\mathcal{M}_{\ell = 1}$/$\mathcal{M}_{\ell = 0}$) to perform the analysis, as shown in Figs \ref{fig:7976303ED} and \ref{fig:7976303mass}. Through the analysis of observed frequencies, we find two modes that depart from the ridge significantly at frequencies of $\sim$ 700 $\mu$Hz and $\sim$ 1000 $\mu$Hz, which means that there are strong mixed-modes. Then, we deduce that the distribution of mode-mass ratios may show two peaks around frequencies of $\sim$ 700 $\mu$Hz and $\sim$ 1000 $\mu$Hz. For models M1, and M6, the mode-mass ratios show no such features, and the frequencies of $\ell$ = 1 modes in the echelle diagrams do not match observations well. For models M4, M5, M7 and M10, although similar behaviors are found in their mode-mass ratios at $\sim$ 700 $\mu$Hz and $\sim$ 1000 $\mu$Hz, in their echelle diagrams the theoretical frequencies significantly deviate from the observations, especially for lower frequencies. We only find three best-fitting models (M8, M9, and M11), which fit all the modes better and have been highlighted in bold font in Table \ref{Tab:7976303}.

From the parameters of the three best-fitting models, we find that there is a difference of $\sim$ 0.03 $M_\odot$ in mass is between the models with [$\alpha$/Fe] = 0.2 and 0.4, while the estimated ages are very close (the difference is about 0.1 Gyr). These three models are regarded better reproducing the observed avoided-crossing behaviours, and point to a star with $M$ = 1.20 $\pm$ 0.02 $M_\odot$, $t$ = 4.88 $\pm$ 0.08 Gyr, $R$ = 2.04 $\pm$ 0.01 $R_\odot$, and $L$ = 5.07 $\pm$ 0.05 $L_\odot$.

\subsubsection{Comparison With Previous Work}

 \begin{table*}
  \centering                          
   \begin{minipage}{140mm}
   \caption[]{Parameters of KIC 7976303 obtained by different models.}
   \label{Tab:compare}
  \begin{tabular}{l c c c c c c c }        
  \hline\hline                 
     KIC     & $\emph{M}$      & $\emph{R}$    & $\emph{L}$    &  $\emph{t}$ & [$\alpha$/Fe]  & Ref. \footnote{RADIUS, YB, SEEK and AMP results come from Marthur et al. (2012)}\\

             & ($M_{\odot}$)  & ($R_{\odot}$) & ($L_{\odot}$) & (Gyr)       &      &   \\  
  \hline                        
      7976303         & 1.04 $\pm$ 0.03          & 1.93 $\pm$ 0.02         & ...  & 5.57 $\pm$ 0.61           &  0.0      & RADIUS     \\      
                      & 1.10 $^{ +~0.05}_{ -~0.08}$  & 2.07 $^{ +~0.05}_{ -~0.07}$ & ...  & 5.65 $^{ +~1.35}_{ -~0.95}$   &  0.0      & YB   \\
                      & 1.05 $^{ +~0.08}_{ -~0.04}$  & 1.98 $^{ +~0.03}_{ -~0.05}$ & ...  & 5.00 $^{ +~0.93}_{ -~0.10}$   &  0.0      & SEEK   \\
                      & 1.17 $\pm$ 0.02          & 2.03 $\pm$ 0.05         & 4.16 & 5.81 $\pm$ 0.03           &  0.0      & AMP   \\
  \hline    
                      & 1.20 $\pm$ 0.02          & 2.04 $\pm$ 0.01         & 5.07 $\pm$ 0.05  & 4.88 $\pm$ 0.08 &  0.2 - 0.4    &  this work \\
  \hline                                   
  \end{tabular}
  \end{minipage}
  \end{table*}

Comparing with the results obtained with the standard models by \citet{Mathur2012}, we obtain a larger mass and a younger age, which can be seen in Table \ref{Tab:compare}. The differences are mainly caused by the effects of $\alpha$-enhancement on the theoretical model. In \citet{Mathur2012}, the characteristics of theoretical models given by AMP method fail to fit observed atmospheric properties and oscillation frequencies simultaneously. Similar results are obtained in this work, and we find no best-fitting model without $\alpha$-enhancement because they can not fit mixed-modes. However, in the other two cases ([$\alpha$/Fe] = 0.2 and 0.4) we obtained models that reproduce all the observational features of the star. This results might be related to the change of opacities, which affects the energy transport in stellar interiors and thus leads to a change in structure.

\begin{figure*}
\centering
\includegraphics[width=16.0cm]{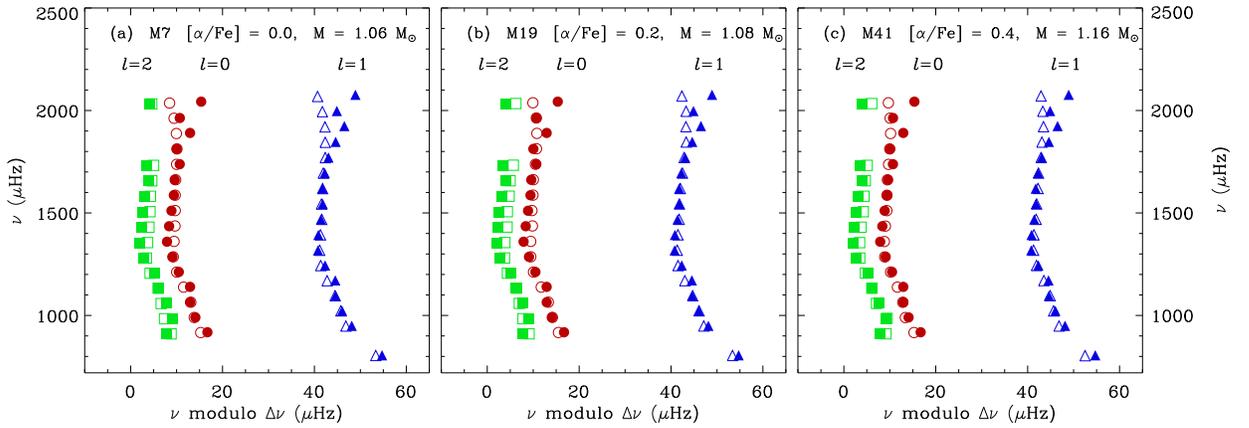}
\caption[]{ \'Echelle diagrams of three best-fitting models (M7, M19, and M41) for KIC 8694723. The filled signs represent observations, the hollow ones represent data from models. Red circles represent frequencies for $\ell$ = 0, blue triangles are frequencies for $\ell$ = 1, green squares correspond to modes for $\ell$ = 2. The values of $\chi_{\nu}^2$ of the models have been presented in the diagrams. The large frequency separation of KIC 8694723 is 74.54 $\nu$ Hz.} 
\label{fig:8694723ED}
\end{figure*}


\subsection{KIC 8694723}

Similarly, for KIC 8694723, we list candidate models ($\chi_{\nu}^2$ $\leq$ 100) for seeking best-fitting models in Table \ref{Tab:8694723all}. To examine the agreement between models and observations, we analyse the behaviours of the oscillation frequencies of the models. From the echelle diagrams, we find that a higher $\chi_{\nu}^2$ value leads to a larger difference between theoretical and observational frequencies. When $\chi_{\nu}^2$ $>$ 20, the models seem to deviate from observations more obviously, either in the high frequencies or in the low frequencies, compared to models with $\chi_{\nu}^2$ $<$ 20. Because the frequencies are very sensitive to the interior structures of the star, the models with $\chi_{\nu}^2$ $<$ 20 are regarded as best-fitting models, which are supposed to be more representative of the stellar structures. The models are highlighted in bold font in Table \ref{Tab:8694723all}. Examples of best-fitting models have been presented in Figure \ref{fig:8694723ED}.


By examining the echelle diagram for each model, we find best-fitting models in all three cases, (i.e., standard models, $\alpha$-enhanced models with [$\alpha$/Fe] = 0.2 and 0.4). Comparing the results of the three cases, models with higher values of [$\alpha$/Fe] favour larger masses and younger ages. This trend is mainly caused by the change in initial metal abundance ($Z_i$), because with a fixed [Fe/H], $\alpha$-enhanced models have a higher value of metal abundance as a result of the change of metal mixtures.

For KIC 8694723, we have found best-fitting models in both scaled-solar and $\alpha$-enhanced cases. These best-fitting models give possible estimations of stellar parameters. The uncertainties of the parameters are likelihood weighted deviations calculated from all the possible models. Standard models point to a star M = 1.05 $\pm$ 0.04 $M_\odot$ and $t$ = 6.26 $\pm$ 0.22 Gyr. For the case of [$\alpha$/Fe] = 0.2, the models point to a star with M = 1.10 $\pm$ 0.03 $M_\odot$ and $t$ = 5.72 $\pm$ 0.30 Gyr. Models with [$\alpha$/Fe] = 0.4 estimate the star as M = 1.17 $\pm$ 0.06 $M_\odot$ and $t$ = 4.94 $\pm$ 0.47 Gyr. Considering the results of the best-fitting models, with an enhancement in the value of [$\alpha$/Fe] by 0.2 dex, the mass increases by $\sim$ 0.08 $M_\odot$, and the age decreases by $\sim$ 1.0 Gyr. Since the star has been observed behaving as $\alpha$-enhanced, and only $\alpha$-enhanced models are used to estimate the stellar parameters, this star might have the characteristics $M$ = 1.13 $\pm$ 0.06 $M_\odot$, $t$ = 5.47 $\pm$ 0.43 Gyr, $R$ = 1.55 $\pm$ 0.13 $R_\odot$ and $L$ = 3.18 $\pm$ 0.26 $L_\odot$.

Compared with KIC 7976303, KIC 8694723 is a post turn-off star, and there are no mixed modes. Because it is not as difficult to fit mixed modes, more models have been selected as best-fitting models and give a probable estimation of the star. Thus, the uncertainties we have obtained are slightly larger for KIC 8694723.

\begin{table*}
 \centering
 \begin{minipage}{160mm}
\caption[]{Theoretical stellar parameters from models for KIC 8694723.}
\label{Tab:8694723all}
 \begin{tabular}{cccccccccccccccc}
 \hline\hline                 
    models& $M$           &	$Z_{i}$ & [$\alpha$/Fe] & $t$   & $L$ & $R$ & $\log$ g & $T_{\rm{eff}}$& $(Z/X)_{s}$ &[Fe/H]   & $\langle\Delta\nu\rangle$ &  $\chi^2_{\nu}$ \\ %
          &($M_{\odot}$)&           &          &  (Gyr) &  ($L_{\odot}$) & ($R_{\odot}$)              &       & (K)        &           &(dex) &      & ($\mu$Hz)      &     \\
     \hline\noalign{\smallskip}
     1 &1.02 & 0.007 & 0.0 & 6.907 & 2.89 &  1.50 &	4.10 &	6154.6 &	0.006 & -0.60    &	74.74 &	45  \\%
     $\textbf{2}$ &$\textbf{1.02}$ & $\textbf{0.007}$ & $\textbf{0.0}$ & $\textbf{6.912}$ & $\textbf{2.90}$ &  $\textbf{1.50}$ &	 $\textbf{4.10}$ & $\textbf{6153.8}$ & $\textbf{0.006}$ & $\textbf{-0.60}$   & $\textbf{74.65}$ &	$\textbf{13}$  \\

     3 &1.04 & 0.007 & 0.0 & 6.308 & 3.02 &  1.50 & 	4.10 &	6207.3 &	0.005 &	-0.63    &	74.76 &	47  \\%

     $\textbf{4}$ &$\textbf{1.04}$ & $\textbf{0.007}$ & $\textbf{0.0}$ & $\textbf{6.313}$ & $\textbf{3.02}$  &  $\textbf{1.51}$ & 	 $\textbf{4.10}$ &	$\textbf{6206.7}$ &	 $\textbf{0.005}$ &	 $\textbf{-0.63}$   &	$\textbf{74.67}$ &	 $\textbf{15}$\\

     5 &1.04 & 0.007 & 0.0 & 6.318 & 3.02 &  1.51 & 	4.10 &	6206.0 &	0.005 &	-0.63    &	74.57 &	55  \\%

6 &1.06 & 0.008 & 0.0 & 6.139 & 3.02 &	1.52 &	4.10 &	6189.0 &	0.007 &	-0.54    &	74.75 &	44 \\%

$\textbf{7}$ &$\textbf{1.06}$ & $\textbf{0.008}$ & $\textbf{0.0}$ & $\textbf{6.144}$ & $\textbf{3.03}$ &	$\textbf{1.52}$ &	 $\textbf{4.10}$ &	$\textbf{6188.3}$ & $\textbf{0.007}$ &	 $\textbf{-0.54}$   & $\textbf{74.65}$ &  	$\textbf{13}$ \\

8 &1.06 & 0.008 & 0.0 & 6.149 & 3.03 &	1.52 &	4.10 &	6187.6 &	0.007 &	-0.54    &	74.56 &	47  \\%

9  & 1.08 & 0.008 & 0.0 & 5.604 & 3.15 &	1.52 &	4.11 &	6239.9 &	0.006 &	-0.57    &	74.79 &	61  \\%
$\textbf{10}$  & $\textbf{1.08}$ & $\textbf{0.008}$  & $\textbf{0.0}$  & $\textbf{5.609}$  & $\textbf{3.15}$ & $\textbf{1.52}$ & $\textbf{4.11}$	 & $\textbf{6239.4}$	 & $\textbf{0.006}$ &	$\textbf{-0.57}$    &	$\textbf{74.69}$ &	$\textbf{16}$  \\%
11  & 1.08 & 0.008 & 0.0 & 5.614 & 3.15 &	1.52 &	4.11 &  6238.8 &	0.006 &	-0.57    &	74.59 &    43   \\%

     \hline\noalign{\smallskip}

12 &1.06 & 0.009 & 0.2 & 6.449 &2.96 &	1.52 &	4.10 &	6145.3 &	0.008 &	-0.61    &	74.74 &	40  \\

$\textbf{13}$ &$\textbf{1.06}$ & $\textbf{0.009}$ & $\textbf{0.2}$ & $\textbf{6.454}$ &$\textbf{2.96}$ &	$\textbf{1.52}$ &	 $\textbf{4.10}$ &	 $\textbf{6144.7}$ &	$\textbf{0.008}$ &	$\textbf{-0.61}$  &	 $\textbf{74.67}$  &  	 $\textbf{14}$\\

14 &1.06 & 0.009 & 0.2 & 6.459 &2.96 &	1.52 &	4.10 &	6144.0 &	0.008 &	-0.61    &	74.53 &	60  \\

15 &1.08 & 0.009 & 0.2 & 5.888 &3.08 &	1.52 &	4.11 &	6197.3 &	0.007 & -0.64    &	74.78 &	59  \\

$\textbf{16}$ &$\textbf{1.08}$ & $\textbf{0.009}$ & $\textbf{0.2}$ & $\textbf{5.893}$ &$\textbf{3.08}$ & $\textbf{1.53}$ & $\textbf{4.11}$ & $\textbf{6196.7}$ & $\textbf{0.007}$ & $\textbf{-0.64}$   & $\textbf{74.64}$ &   $\textbf{11}$  \\

17 &1.08 & 0.009 & 0.2 & 5.898 &3.09 &	1.53 &	4.10 &	6196.1 &	0.007 &	-0.64    &	74.63 &	26  \\%

18 &1.08 & 0.010 & 0.2 & 6.213 &2.99 &	1.53 &	4.10 &	6142.6 &	0.009 &	-0.56    &	74.72 &	37  \\%

$\textbf{19}$ &$\textbf{1.08}$ & $\textbf{0.010}$ &$\textbf{0.2}$ & $\textbf{6.218}$ &$\textbf{2.99}$ &$\textbf{1.53}$ & $\textbf{4.10}$ &	 $\textbf{6141.9}$ &$\textbf{0.009}$ &$\textbf{-0.56}$   & $\textbf{74.66}$ &    $\textbf{11}$  \\

20 &1.08 & 0.010 & 0.2 & 6.223 &2.99 &	1.53 &	4.10 &	6141.3 &	0.009 &	-0.56   &	74.57 &	33  \\%

21 &1.10 & 0.010 & 0.2 & 5.666 &3.11 &	1.53 &	4.11 &	6193.7 &	0.008 &	-0.59   &	74.75 &	41  \\%

$\textbf{22}$ &$\textbf{1.10}$ & $\textbf{0.010}$ & $\textbf{0.2}$ & $\textbf{5.671}$ &$\textbf{3.11}$ & $\textbf{1.54}$ &$\textbf{4.11}$ &$\textbf{6193.1}$ & $\textbf{0.008}$ & $\textbf{-0.59}$  & $\textbf{74.64}$ &  $\textbf{13}$  \\

23 &1.10 & 0.010 & 0.2 & 5.676 &3.12 &	1.54 &	4.11 &	6192.6 &	0.008 &	-0.59    &	74.57 &	47  \\%

24 &1.12 & 0.010 & 0.2 & 5.168 &3.23 &	1.54 &	4.11 &	6244.0 &	0.008 &	-0.62    &	74.77 &	47  \\%

$\textbf{25}$ & $\textbf{1.12}$ & $\textbf{0.010}$ & $\textbf{0.2}$ & $\textbf{5.173}$ &$\textbf{3.24}$ & $\textbf{1.54}$ & $\textbf{4.11}$ & $\textbf{6243.4}$ & $\textbf{0.008}$ & $\textbf{-0.62}$  & $\textbf{74.68}$ &  $\textbf{15}$  \\

26 &1.12 & 0.010 & 0.2 & 5.178 &3.24 &	1.54 &	4.11 &	6242.8 &	0.008 &	-0.62    &	74.57 &	51  \\%

27 &1.12 & 0.011 & 0.2 & 5.460 &3.14 &	1.54 &	4.11 &	6190.4 &	0.009 &	-0.54    &	74.68 &	23  \\%
28 &1.12 & 0.011 & 0.2 & 5.465 &3.14 &	1.54 &	4.11 &	6189.9 &	0.009 &	-0.54    &	74.75 &	46  \\%

$\textbf{29}$ & $\textbf{1.12}$ & $\textbf{0.011}$ & $\textbf{0.2}$ & $\textbf{5.470}$ &$\textbf{3.14}$ & $\textbf{1.54}$ & $\textbf{4.11}$ & $\textbf{6189.3}$ & $\textbf{0.009}$ & $\textbf{-0.54}$    & $\textbf{74.66}$ &   $\textbf{12}$  \\

30  &  1.12  &  0.011  &  0.2 &  5.475  & 3.15  &  1.55  &  4.11  &  6188.7  &  0.009  &  -0.54   &  74.55 &   48    \\%

31 &1.14 & 0.011 & 0.2 & 4.986 &3.26 &	1.55 &	4.12 &	6239.4 &	0.009 &	-0.57    &	74.76 &	44  \\%

$\textbf{32}$ & $\textbf{1.14}$ & $\textbf{0.011}$ & $\textbf{0.2}$ & $\textbf{4.991}$ &$\textbf{3.27}$ & $\textbf{1.55}$ & $\textbf{4.11}$ & $\textbf{6238.8}$ & $\textbf{0.009}$ & $\textbf{-0.57}$    & $\textbf{74.68}$ &   $\textbf{15}$   \\

33 &1.14 & 0.011 & 0.2 & 4.996 &3.27 &	1.55 &	4.11 &	6238.2 &	0.009 &	-0.57   &	74.59 &	43  \\%

      \hline\noalign{\smallskip}

34 &1.14 & 0.013 & 0.4 & 5.397 &3.11 &	1.56 &	4.11 &	6151.5 &	0.012 & -0.61	 &	74.72 &	37  \\%

$\textbf{35}$ &$\textbf{1.14}$ & $\textbf{0.013}$ & $\textbf{0.4}$ & $\textbf{5.402}$ &$\textbf{3.11}$ &	$\textbf{1.56}$ & $\textbf{4.11}$ & $\textbf{6150.9}$ &	$\textbf{0.012}$ & $\textbf{-0.61}$    &	 $\textbf{74.63}$ &   $\textbf{9}$  \\

36 &1.14 & 0.013 & 0.4 & 5.407 &3.12 &	1.56 &	4.11 &	6150.4 &	0.012 & -0.61	  &	74.55 &	37  \\%

37 &1.16 & 0.013 & 0.4 & 4.923 &3.23 &	1.56 &	4.12 &	6202.3 &	0.011 & -0.63	  &	74.75 &	47  \\%

$\textbf{38}$ &$\textbf{1.16}$ & $\textbf{0.013}$ & $\textbf{0.4}$ & $\textbf{4.928}$ &$\textbf{3.24}$ &	$\textbf{1.56}$ & $\textbf{4.12}$ & $\textbf{6201.7}$ & $\textbf{0.011}$ & $\textbf{-0.63}$    &	 $\textbf{74.65}$ &  $\textbf{12}$  \\

39 &1.16 & 0.013 & 0.4 & 4.933 &3.24 &	1.56 &	4.12 &	6201.1 &	0.011 & -0.63	  &	74.57 &	33  \\%

40 &1.16 & 0.014 & 0.4 & 5.156 &3.15 &	1.56 &	4.11 &	6154.6 &	0.013 & -0.57	   &	74.73 &	41  \\%

$\textbf{41}$ &$\textbf{1.16}$ & $\textbf{0.014}$ & $\textbf{0.4}$ & $\textbf{5.161}$ &$\textbf{3.15}$ &	$\textbf{1.56}$ & $\textbf{4.11}$ & $\textbf{6154.0}$ & $\textbf{0.013}$ & $\textbf{-0.57}$   &	 $\textbf{74.63}$ &  $\textbf{9}$  \\

42 &1.16 & 0.014 & 0.4 & 5.166 &3.16 &	1.57 &	4.11 &	6153.5 &	0.013 & -0.57	   &	74.55 &	35  \\%

43 &1.18 & 0.014 & 0.4 & 4.698 &3.27 &	1.57 &	4.12 &	6205.3 &	0.012 & -0.59     &	74.77 &	81  \\

44 &1.18 & 0.014 & 0.4 & 4.703 &3.28 &	1.57 &	4.12 &	6204.7 &	0.012 & -0.59     &	74.73 &	37  \\

$\textbf{45}$ &$\textbf{1.18}$ & $\textbf{0.014}$ & $\textbf{0.4}$ & $\textbf{4.708}$ &$\textbf{3.28}$ &	$\textbf{1.57}$ & $\textbf{4.12}$ & $\textbf{6204.1}$ & $\textbf{0.012}$ & $\textbf{-0.59}$   &	 $\textbf{74.64}$ &   $\textbf{12}$  \\

46 &1.18 & 0.014 & 0.4 & 4.713 &3.28 &	1.57 &	4.12 &	6203.6 &	0.012 & -0.59      &	74.56 &	41  \\

47 &1.18 & 0.015 & 0.4 & 4.913 & 3.20 &  1.57  &	 4.12  &  6160.4  &  0.014  & -0.54     &	74.72 &	43   \\

$\textbf{48}$ &$\textbf{1.18}$ & $\textbf{0.015}$ & $\textbf{0.4}$ & $\textbf{4.918}$ &$\textbf{3.20}$ &	$\textbf{1.57}$ & $\textbf{4.12}$ & $\textbf{6159.9}$ &	$\textbf{0.014}$ & $\textbf{-0.54}$   &	 $\textbf{74.64}$ &   $\textbf{10}$  \\

49 &1.18 & 0.015 & 0.4 & 4.923 & 3.20 &  1.57  &	 4.12  &  6159.4  &  0.014  &  -0.54    &  74.56 &   29   \\

50 &1.18 & 0.015 & 0.4 & 4.928 &3.21 &	1.58 &	4.12 &	6158.9 &	0.014 & -0.54      &	74.49 &	90  \\

51 &1.20 & 0.014 & 0.4 & 4.274 &3.40 &	1.57 &	4.12 &	6253.5 &	0.011 & -0.62     &	74.75 &	43  \\

$\textbf{52}$ &$\textbf{1.20}$ & $\textbf{0.014}$ & $\textbf{0.4}$ & $\textbf{4.279}$ &$\textbf{3.40}$ &	$\textbf{1.58}$ & $\textbf{4.12}$ & $\textbf{6253.0}$ & $\textbf{0.011}$ & $\textbf{-0.62}$    &	 $\textbf{74.67}$ &   $\textbf{16}$  \\

53 &1.20 & 0.014 & 0.4 & 4.284 &3.41 &	1.58 &	4.12 &	6252.6 &	0.011 & -0.62     &	74.58 &	38  \\

54 &1.20 & 0.015 & 0.4 & 4.471 &3.32 &	1.58 &	4.12 &	6209.8 &	0.013 & -0.56     &	74.75 &	51  \\

$\textbf{55}$ &$\textbf{1.20}$ & $\textbf{0.015}$ & $\textbf{0.4}$ & $\textbf{4.476}$ &$\textbf{3.33}$ &	$\textbf{1.58}$ & $\textbf{4.12}$ & $\textbf{6209.4}$ & $\textbf{0.013}$ & $\textbf{-0.56}$    & $\textbf{74.66}$ &   $\textbf{13}$  \\

56 &1.20 & 0.015 & 0.4 & 4.481 &3.33 &	1.58 &	4.12 &	6208.8 &	0.013 & -0.56     &	74.58 &	30  \\
57 &1.20 & 0.015 & 0.4 & 4.486 &3.33 &	1.58 &	4.12 &	6208.4 &	0.013 & -0.56     &	74.50 &	95  \\

58 &1.22 & 0.015 & 0.4 & 4.049 & 3.48 &	1.58 &	4.12 &	6269.9 &	0.012 & -0.58     &	74.53 &	74  \\

 \hline               
\end{tabular}
\end{minipage}
\end{table*}

\section{ Discussions and Conclusions}

In this work, theoretical modelling and asteroseismic analysis are carried out to determine the evolution status and stellar parameters of two $\alpha$-enhanced stars, KIC 7976303 and KIC 8694723. To achieve better estimations, the effects of $\alpha$-enhancement are considered in the stellar models.

KIC 7976303 is an evolved subgiant star with mixed modes. Models without $\alpha$-enhancement for KIC 7976303 fit the observations well, except for the behaviours of mixed modes, while several $\alpha$-enhanced models succeed in fitting all features. This result could be caused by the influence in the interior of the change of opacities, but could also be caused by the free parameters in the theoretical model, such as the initial helium abundance. The results of this work show that $\alpha$-enhanced models present better agreements with the observed data than the standard models. These models indicate that KIC 7976303 is a star with $M$ = 1.20 $\pm$ 0.02 $M_\odot$, $t$ = 4.88 $\pm$ 0.08 Gyr, $R$ = 2.04 $\pm$ 0.01 $R_\odot$, and $L$ = 5.07 $\pm$ 0.05 $L_\odot$

Considering our analysis, KIC 8694723 is a turn-off star. We find best-fitting models in all three cases ([$\alpha$/Fe] = 0.0, 0.2, 0.4). The models with $\alpha$-enhancement estimate larger mass and younger age than standard models. According to the observed metal abundances of this star, we suggest using $\alpha$-enhanced models to estimate the stellar parameters, which are $M$ = 1.13 $\pm$ 0.06 $M_\odot$, $t$ = 5.47 $\pm$ 0.43 Gyr, $R$ = 1.55 $\pm$ 0.13 $R_\odot$, and $L$ = 3.18 $\pm$ 0.26 $L_\odot$. With more precise observations in the future, we might be able to estimate the fundamental properties of this star more accurately.

From the analysis of these two stars, we find that for a certain star, models with scaled-solar and $\alpha$-enhanced mixtures determine significantly different masses and ages. The results presented by best-fitting models indicate that the uncertainty of metal mixture will cause obvious bias on our estimations of a star. These bias between models suggest that, in the study of stars that might be $\alpha$-enhanced, such as metal-poor field stars and members of GCs, high-quality spectroscopic observations are of much importance, while $\alpha$-enhancement should be included in modelling processes to eliminate possible errors.

\section*{Acknowledgments}

Funding for this Discovery mission is provided by NASA¡¯s Science Mission Directorate. We would like to express the sincere gratitude to D. A. VandenBerg, J. W. Ferguson, Y. -C. Kim, T. R. Bedding, and T. Appourchaux for their generous helps. We also thank Y. Q. Chen for her advices. This work is supported by grants 11273007, 11273012 and 10933002
from the National Natural Science Foundation of China, and the
Fundamental Research Funds for the Central Universities.

\appendix

\section[]{Calculation of [$\alpha$/Fe] from metal mixtures of the stars}

\begin{table}
 \begin{minipage}{140mm}
\caption[]{metal mixture for KIC 7976303.\footnote{$\log$ N$_H$ = 12.00}}
\label{Tab:7976303mix}
 \begin{tabular}{cccccccccccccccc}
 \hline\hline                 
    element & $\log$ N$_\odot$     &	enhancement  &    $\log$ N$_{star}$       	\\
       \hline\noalign{\smallskip}
     C      & 8.52    &  0.13  &  8.65        \\%
     N      & 7.92    &  0.00  &  7.92        \\%
     O      & 8.83    &  0.43  &  9.26        \\%
     F      & 4.56    &  0.00  &  4.56        \\%
     Ne     & 8.08    &  0.00  &  8.08        \\%
     Na     & 6.33    &  0.04  &  6.37        \\%
     Mg     & 7.58    &  0.22  &  7.80        \\%
     Al     & 6.47    &  0.00  &  6.47    	  \\%
     Si     & 7.55    &  0.08  &  7.63        \\%
     P      & 5.45    &  0.00  &  5.45        \\%
     S      & 7.33    &  0.00  &  7.33        \\%
     Cl     & 5.50    &  0.00  &  5.50        \\%
     Ar     & 6.40    &  0.00  &  6.40        \\%
     K      & 5.12    &  0.00  &  5.12        \\%
     Ca     & 6.36    &  0.11  &  6.47        \\%
     Sc     & 3.17    &  0.00  &  3.17        \\%
     Ti     & 5.02    &  0.11  &  5.13        \\%
     V      & 4.00    &  0.15  &  4.15        \\%
     Cr     & 5.67    & -0.09  &  5.58        \\%
     Mn     & 5.39    &  0.00  &  5.39        \\%
     Fe     & 7.50    &  0.00  &  7.50        \\%
     Co     & 4.92    &  0.00  &  4.92        \\%
     Ni     & 6.25    & -0.06  &  6.19        \\%
 \hline               
\end{tabular}
\end{minipage}
\end{table}

\begin{table}
 \begin{minipage}{140mm}
\caption[]{metal mixture for KIC 8694723.\footnote{$\log$ N$_H$ = 12.00}}
\label{Tab:8694723mix}
 \begin{tabular}{cccccccccccccccc}
 \hline\hline                 
    element & $\log$ N$_\odot$     &	enhancement  &    $\log$ N$_{star}$    	\\
       \hline\noalign{\smallskip}
     C      & 8.52    &  0.20  &  8.72         \\%
     N      & 7.92    &  0.21  &  8.13        \\%
     O      & 8.83    &  0.41  &  9.24        \\%
     F      & 4.56    &  0.00  &  4.56        \\%
     Ne     & 8.08    &  0.00  &  8.08        \\%
     Na     & 6.33    &  0.09  &  6.42        \\%
     Mg     & 7.58    &  0.22  &  7.80        \\%
     Al     & 6.47    &  0.00  &  6.47    	  \\%
     Si     & 7.55    &  0.14  &  7.69        \\%
     P      & 5.45    &  0.00  &  5.45        \\%
     S      & 7.33    &  0.00  &  7.44        \\%
     Cl     & 5.50    &  0.00  &  5.50        \\%
     Ar     & 6.40    &  0.00  &  6.40        \\%
     K      & 5.12    &  0.00  &  5.12        \\%
     Ca     & 6.36    &  0.04  &  6.40        \\%
     Sc     & 3.17    &  0.00  &  3.17        \\%
     Ti     & 5.02    &  0.02  &  5.04        \\%
     V      & 4.00    &  0.25  &  4.25        \\%
     Cr     & 5.67    & -0.05  &  5.62        \\%
     Mn     & 5.39    &  0.00  &  5.39        \\%
     Fe     & 7.50    &  0.00  &  7.50        \\%
     Co     & 4.92    &  0.00  &  4.92        \\%
     Ni     & 6.25    & -0.06  &  6.19        \\%
 \hline               
\end{tabular}
\end{minipage}
\end{table}

\citet{Bruntt2012} presented element abundances for KIC 7976303 and KIC 8694723 in their Table 4. The complete version of this table is from on-line data\footnote{http://vizier.cfa.harvard.edu/viz-bin/VizieR?-source=J/MNRAS/423/122}.

For scaled-solar metal element mixture, the elements have the same proportion with solar metal elements, [M/Fe] = [M/Fe]$_\odot$ = 0.0 (M corresponds to metal element). For special metal mixtures, such as $\alpha$-enhanced ones, [$\alpha$/Fe] $>$ 0.0 ($\alpha$ corresponds to $\alpha$-element). The enhancement of a single element is considered as
\begin{equation}
\left[ {M/Fe} \right] = \log \left( {\frac{{N_M }}{{N_{Fe} }}} \right)_{star}  - \log \left( {\frac{{N_M }}{{N_{Fe} }}} \right)_ \odot   = \log \left( {N_M } \right)_{star}  - \log \left( {N_M } \right)_ \odot  .
\end{equation}
Where N stands for the number of the particles in a unit volume, i.e. the abundance by number. From this relation we consider [M/Fe] as the enhancement of a metal element to the solar-mixture. The value of [M/Fe] can be calculated from the observed element abundances [M/Fe] = [M/H]-[Fe/H]. With the value of [M/Fe] we can construct a metal mixture for this star. We use the \citet{GS98} scaled-solar mixture to start on the system in which $\log$ $N_{\rm{H}}$ = 12.0. Tables \ref{Tab:7976303mix} and Table \ref{Tab:8694723mix} present the metal mixture of KIC 7976303 and KIC 8694723.

The value of [$\alpha$/Fe] means the average enhancement of $\alpha$-elements ($\alpha$-elements in this work include O, Ne, Mg, Si, S, Ca, Ti, Ar was ignored). The value of [$\alpha$/Fe] can be estimated as
\begin{equation}
[\alpha /Fe] = \log \left( {\frac{{\sum {N_\alpha  } }}{{N_{Fe} }}} \right)_{star}  - \log \left( {\frac{{\sum {N_\alpha  } }}{{N_{Fe} }}} \right)_ \odot   = \log \left[ {\frac{{\left( {\sum {N_\alpha  } } \right)_{star} }}{{\left( {\sum {N_{\alpha}  } } \right)_ \odot  }}} \right].
\end{equation}
Using the $\log$ N values we can obtain the value of [$\alpha$/Fe] for each star.

\section[]{Calculation of Z/X from metal mixtures}

For Population I stars, the relationship of [Fe/H] with ratio of surface heavy-element abundance to hydrogen abundance ($Z/X$) is $\log$ ($Z/X$) = $\log$ ($Z/X$)$_\odot$ + [Fe/H], where ($Z/X$)$_\odot$ is the ratio of the heavy element to hydrogen for scaled-solar mixture. While this relationship is not adequate for $\alpha$-enhanced mixtures. For a certain [Fe/H], an $\alpha$-enhanced mixture leads to a larger $Z/X$ than scaled-solar mixture, we use detailed element abundance mixtures to calculate $Z/X$.

The metal mixture we have used in opacity tables and in the models is with [$\alpha$/Fe] = 0.2 (0.4), which means adding 0.2 (0.4) dex to the solar $\log$ $N$ values for O ,Ne, Mg, Si, S, Ca, Ti. These mixtures start from the \citet{GS98} solar metal-mixture. Here, we introduce two calculation methods to obtain the mass fraction ratio Z/X.

To simplify the description, we take [Fe/H] = 0.0 for example. In the following description, $i$ represent all the elements, including Hydrogen and Helium, $M$ represent metal elements only.

\subsection[]{Calculation Method 1}

We convert the $\log$ $N_i$ values to $N_i$ (i.e., take the antilog; $i$ represents all the elements including H, He). We sum the $N_i$, and then calculate $ A_i  =\frac{{N_i }}{{\sum\limits_i {N_i } }}$, and we obtain the number-fraction abundances for each element $i$ (we call this A$_i$). To obtain the equivalent mass-fraction abundances, $X_i$, we calculate the sum of ($A_i\cdot W_i$), where $W_i$ is the atomic weight of element species $i$, and then we calculate $X_i  = \frac{{A_i W_i }}{{\sum\limits_i {A_i W_i } }}$.

From the definition of the mass fraction, we describe the mass abundance of all the metal elements as follows:
\begin{equation}
Z = X_{metal}  = \frac{{\sum\limits_M {A_M W_M } }}{{\sum\limits_i {A_i W_i } }}.
  \label{Z}
\end{equation}
Then, Z/X can be calculated as,
\begin{equation}
Z/X = \frac{{\sum\limits_M {A_M W_M } }}{{A_H W_H }}.
  \label{ZX}
\end{equation}

Because the number fraction is defined as $ A_i  =\frac{{N_i }}{{\sum\limits_i {N_i } }}$, we can expand Equation (\ref{ZX}) as follows,

\begin{equation}
Z/X = \frac{{\sum\limits_M {A_M W_M } }}{{A_H W_H }} = \frac{{\sum\limits_M {\frac{{N_M W_M }}{{\sum\limits_i {N_i } }}} }}{{\frac{{N_H W_H }}{{\sum\limits_i {N_i } }}}} = \frac{{\sum\limits_M {N_M W_M } }}{{N_H W_H }}
  \label{ZX2}
\end{equation}.

We use the $\log$ $N_{\rm{H}}$ = 12.0 system, so that $N_{\rm{H}}$ = $10^{12}$. With the metal mixtures presented in Tables \ref{Tab:a02mix} and \ref{Tab:a04mix}, we can calculate the value of $Z/X$ without the value of $N_{\rm{He}}$.

Furthermore, if we know the value of helium abundance, $Y$, we can use following equation to obtain $N_{\rm{He}}$,

\begin{equation}
\begin{array}{l}
 \frac{Y}{{1 - Y}} = \frac{Y}{{X + Z}} = \frac{{\frac{{A_{He} W_{He} }}{{\sum\limits_i {A_i W_i } }}}}{{\frac{{\sum\limits_M {A_M W_M } }}{{\sum\limits_i {A_i W_i } }} + \frac{{A_H W_H }}{{\sum\limits_i {A_i W_i } }}}} = \frac{{A_{He} W_{He} }}{{\sum\limits_M {A_M W_M }  + A_H W_H }} \\
  = \frac{{\frac{{N_{He} }}{{\sum\limits_i {N_i } }}W_{He} }}{{\frac{{\sum\limits_M {N_M } W_M }}{{\sum\limits_i {N_i } }} + \frac{{N_H }}{{\sum\limits_i {N_i } }}W_H }} = \frac{{N_{He} W_{He} }}{{\sum\limits_M {N_M W_M }  + N_H W_H }}, \\
 \end{array}
 \label{Nhe}
\end{equation}
Thus,
\begin{equation}
N_{He}  = \frac{Y}{{1 - Y}} \cdot \frac{{\sum\limits_M {N_M W_M }  + N_H W_H }}{{W_{He} }}.
  \label{Nhe2}
\end{equation}

\subsection[]{Calculation Method 2}

To confirm the results we obtained by Method 1, we use another method, the iterate process presented by \citet{VandenBerg2014}, to calculate metal abundance, $Z$.

We begin with the helium abundance, $Y$,

\begin{equation}
Y = X_{He}  = \frac{{A_{He} W_{He} }}{{\sum\limits_i {A_i W_i } }}.
  \label{Y}
\end{equation}
Because we have known the mass abundance of helium, $Y$ = 0.248, we can calculate $N_{\rm{He}}$ by iterating. Assuming N$_{He}$ = 10.9, for example, we use Equation (\ref{Y}) to calculate $Y$. We repeat this process until $X_{He}$ is equal to the desired value of $Y$ (e.g. $Y$ = 0.248), and then we determine the value of $N_{\rm{He}}$.
Using the following equation,
\begin{equation}
X = X_H  = \frac{{A_H W_H }}{{\sum\limits_i {A_i W_i } }}
  \label{X}
\end{equation}
we can then calculate $X$, so that $Z$ = 1 - $X$ - $Y$. The results obtained by the two methods are the same.

For a certain star, taking [Fe/H] = -0.5 for example, we adjust all of the $\log$ $N$ abundances by -0.5; that is, $\log$ $N_{M}'$ = $\log$ $N_M$+[Fe/H] ($M$ represents the metal element). Using the new $\log$ $N_M'$, we repeat the calculation processes mentioned upward, then we could obtain the value of $Z/X$ for a certain [Fe/H].

\begin{table*}

 \begin{minipage}{140mm}
 \centering
\caption[]{metal mixture for [$\alpha$/Fe] = 0.2. All the $\alpha$-elements (O, Ne, Mg, Si, S, Ca, Ti; except Ar) have been increased by 0.2 dex.}
\label{Tab:a02mix}
 \begin{tabular}{cccccccccccccccc}
 \hline\hline                 
    element & $\log$ N$_\odot$     &	enhancement  &    $\log$ N$_{\alpha}$   &  Weigh      	 \\
       \hline\noalign{\smallskip}
     C      & 8.52    &  0.00  &  8.52  &  12.01     \\%
     N      & 7.92    &  0.00  &  7.92  &  14.01     \\%
     O      & 8.83    &  0.20  &  9.03  &  16.00     \\%
     F      & 4.56    &  0.00  &  4.56  &  19.00     \\%
     Ne     & 8.08    &  0.20  &  8.28  &  20.18     \\%
     Na     & 6.33    &  0.00  &  6.33  &  22.99     \\%
     Mg     & 7.58    &  0.20  &  7.78  &  24.31     \\%
     Al     & 6.47    &  0.00  &  6.47  &  26.98  	 \\%
     Si     & 7.55    &  0.20  &  7.75  &  28.09     \\%
     P      & 5.45    &  0.00  &  5.45  &  30.97     \\%
     S      & 7.33    &  0.20  &  7.53  &  32.07     \\%
     Cl     & 5.50    &  0.00  &  5.50  &  35.45     \\%
     Ar     & 6.40    &  0.00  &  6.40  &  39.95     \\%
     K      & 5.12    &  0.00  &  5.12  &  39.10     \\%
     Ca     & 6.36    &  0.20  &  6.56  &  40.08     \\%
     Sc     & 3.17    &  0.00  &  3.17  &  44.96     \\%
     Ti     & 5.02    &  0.20  &  5.22  &  47.87     \\%
     V      & 4.00    &  0.00  &  4.00  &  50.94     \\%
     Cr     & 5.67    &  0.00  &  5.67  &  52.00     \\%
     Mn     & 5.39    &  0.00  &  5.39  &  54.94     \\%
     Fe     & 7.50    &  0.00  &  7.50  &  55.85     \\%
     Co     & 4.92    &  0.00  &  4.92  &  58.93     \\%
     Ni     & 6.25    &  0.00  &  6.25  &  58.69     \\%
 \hline               
\end{tabular}
\end{minipage}
\end{table*}

\begin{table*}
 \begin{minipage}{140mm}
  \centering
\caption[]{metal mixture for [$\alpha$/Fe] = 0.4. All the $\alpha$-elements (O, Ne, Mg, Si, S, Ca, Ti; except Ar) have been increased by 0.4 dex.}
\label{Tab:a04mix}
 \begin{tabular}{cccccccccccccccc}
 \hline\hline                 
    element & $\log$ N$_\odot$     &	enhancement  &    $\log$ N$_{\alpha}$   &  Weigh      	 \\
       \hline\noalign{\smallskip}
     C      & 8.52    &  0.00  &  8.52  &  12.01     \\%
     N      & 7.92    &  0.00  &  7.92  &  14.01     \\%
     O      & 8.83    &  0.40  &  9.23  &  16.00     \\%
     F      & 4.56    &  0.00  &  4.56  &  19.00     \\%
     Ne     & 8.08    &  0.40  &  8.48  &  20.18     \\%
     Na     & 6.33    &  0.00  &  6.33  &  22.99     \\%
     Mg     & 7.58    &  0.40  &  7.98  &  24.31     \\%
     Al     & 6.47    &  0.00  &  6.47  &  26.98  	 \\%
     Si     & 7.55    &  0.40  &  7.95  &  28.09     \\%
     P      & 5.45    &  0.00  &  5.45  &  30.97     \\%
     S      & 7.33    &  0.40  &  7.73  &  32.07     \\%
     Cl     & 5.50    &  0.00  &  5.50  &  35.45     \\%
     Ar     & 6.40    &  0.00  &  6.40  &  39.95     \\%
     K      & 5.12    &  0.00  &  5.12  &  39.10     \\%
     Ca     & 6.36    &  0.40  &  6.76  &  40.08     \\%
     Sc     & 3.17    &  0.00  &  3.17  &  44.96     \\%
     Ti     & 5.02    &  0.40  &  5.42  &  47.87     \\%
     V      & 4.00    &  0.00  &  4.00  &  50.94     \\%
     Cr     & 5.67    &  0.00  &  5.67  &  52.00     \\%
     Mn     & 5.39    &  0.00  &  5.39  &  54.94     \\%
     Fe     & 7.50    &  0.00  &  7.50  &  55.85     \\%
     Co     & 4.92    &  0.00  &  4.92  &  58.93     \\%
     Ni     & 6.25    &  0.00  &  6.25  &  58.69     \\%
 \hline               
\end{tabular}
\end{minipage}
\end{table*}

\bsp

\label{lastpage}

\end{document}